\documentclass{amsproc}
\usepackage[utf8]{inputenc}
\usepackage[english]{babel}
\usepackage{booktabs}
\usepackage[margin=2cm]{geometry}
\usepackage{tikz}
\usepackage{multicol}

\newcommand{\ra}[1]{\renewcommand{\arraystretch}{#1}}

\begin{document}
\title{Neural network approach to classifying alarming student responses to online assessment}
\author{Christopher M. Ormerod}
\address{Christopher M. Ormerod 1000 Thomas Jefferson St NW \#200, Washington, DC 20007}
\curraddr{}
\email{cormerod@air.org}

\author{Amy E. Harris}
\address{Amy E. Harris 1000 Thomas Jefferson St NW \#200, Washington, DC 20007}
\curraddr{}
\email{aharris@air.org}

\subjclass[2010]{Primary }


\begin{abstract}
Automated scoring engines are increasingly being used to score the free-form text responses that students give to questions. Such engines are not designed to appropriately deal with responses that a human reader would find alarming such as those that indicate an intention to self-harm or harm others, responses that allude to drug abuse or sexual abuse or any response that would elicit concern for the student writing the response. Our neural network models have been designed to help identify these anomalous responses from a large collection of typical responses that students give. The responses identified by the neural network can be assessed for urgency, severity and validity more quickly by a team of reviewers than otherwise possible. Given the anomalous nature of these types of responses, our goal is to maximize the chance of flagging these responses for review given the constraint that only a fixed percentage of responses can viably be assessed by a team of reviewers.
\end{abstract}

\maketitle

\section{Introduction}

Automated Essay Scoring (AES) and Automated Short Answer Scoring (ASAS) has become more prevalent among testing agencies \cite{ETS, IEA, PEG,Intellimetric}. These systems are often designed to address one task and one task alone; to determine whether a written piece of text addresses a question or not. These engines were originally based on either hand-crafted features or term frequency–inverse document frequency (TF-IDF) approaches \cite{BOW}. More recently, these techniques have been superseded by the combination of word-embeddings and neural networks \cite{Embedding, NNAES1, NNAES2}. For semantically simple responses, the accuracy of these approaches can often be greater than accuracy of human raters, however, these systems are not trained to appropriately deal with the anomalous cases in which a student writes something that elicits concern for the writer or those around them, which we simply call an `alert'. Typically essay scoring systems do not handle alerts, but rather, separate systems must be designed to process these types of responses before they are sent to the essay scoring system. Our goal is not to produce a classification, but rather to use the same methods developed in AES, ASAS and sentiment analysis \cite{TangSentiment, WangSentiment} to identify some percentage of responses that fit patterns seen in known alerts and send them to be assessed by a team of reviewers.

Assessment organizations typically perform some sort of alert detection as part of doing business. In among hundreds of millions of long and short responses we find cases of alerts in which students have outlined cases of physical abuse, drug abuse, depression, anxiety, threats to others or plans to harm themselves \cite{EssayHelp}. Such cases are interesting from a linguistic, educational, statistical and psychological viewpoint \cite{Language}. While some of these responses require urgent attention, given the volume of responses many testing agencies deal with, it is not feasible to systematically review every single student response within a reasonable time-frame. The benefits of an automated system for alert detection is that we can prioritize a small percentage which can be reviewed quickly so that clients can receive alerts within some fixed time period, which is typically 24 hours. Given the prevalence of school shootings and similarly urgent situations, reducing the number of false positives can effectively speed up the review process and hence optimize our clients ability to intervene when necessary.  

As a classification problem in data science, our problem has all the hallmarks of the most difficult problems in natural language processing (NLP) \cite{LillyNLP}; alerts are anomalous in nature making training difficult, the data is messy in that it contains misspellings (both misused real words and incorrectly spelled words) \cite{Spelling}, students often use student specific language or multi-word colloquialisms \cite{Multiword} and the semantics of alerts can be quite complex and subtle, especially when the disturbing content is implicit rather than explicit. The responses themselves are drawn from a wide range of free-form text responses to questions and student comments from a semantically diverse range of topics, including many that are emotive in nature. For example, the semantic differences between an essay on gun-control and a student talking about getting a gun can be very subtle. Sometimes our systems include essays on emotive topics because the difference in language between such essays and alerts can be very small. Students often use phrases like ``kill me now" as hyperbole out of frustration rather than a genuine desire to end ones life, e.g., "this test is so boring, kill me now". To minimize false positives, the engine should attempt to evaluate context, not just operate on key words or phrases.    

When it comes to neural network design, there are two dominant types of neural networks in NLP; convolutional neural networks (CNN) and recurrent neural networks (RNN)\cite{RNNNLP}. Since responses may be of an arbitrary length different recurrent neural networks are more appropriate tools for classifying alerts \cite{CNNNLP}. The most common types of cells used in the design of recurrent neural networks are Gated Recurrent Units (GRU)s \cite{GRU2014} and Long-Short-Term-Memory (LSTM) units \cite{LSTM1997}. The latter were originally designed to overcome the vanishing gradient problem \cite{VanishingGrad}. The GRU has some interesting properties which simplify the LSTM unit and the two types of units can give very similar results \cite{GRUvsLSTM}. We also consider stacked versions, bidirectional variants \cite{Bidirectional1997} and the effect of an attention mechanism \cite{Bahdanau:Attention}. This study has been designed to guide the creation of our desired final production model, which may include higher stacking, dropouts (both regular and recurrent) and may be an ensemble of various networks tuned to different types of responses \cite{Ensemble}. Similar comparisons of architectures have appeared in the literature \cite{Tweets, NNAES2}, however, we were not able to find similar comparisons for detecting anomalous events.

In section \ref{Data} we outline the nature of the data we have collected, a precise definition of an alert and how we processed the data for the neural network. In section \ref{RNNs} we outline the definition of the models we evaluate and how they are defined. In section \ref{sec:Results} we outline our methodology in determining which models perform best given representative sensitivities of the engine. We attempt to give an approximation of the importance of each feature of the final model. 

\section{Defining the Data}\label{Data}

The American Institutes for Research tests up to 1.8 million students a day during peak testing periods. Over the 2016--2017 period AIR delivered 48 million online tests across America. Each test could involve a number of comments, notes and long answer free-form text responses that are considered to be a possible alerts as well as equations or other interactive items that are not considered to be possible alerts. In a single year we evaluate approximately 90 million free-form text responses which range anywhere from a single word or number to ten thousand word essays. These responses are recorded in html and embedded within an xml file along with additional information that allows our clients to identify which student wrote the response. The first step in processing such a response is to remove tags, html code and any non-text using regular expressions. 

To account for spelling mistakes, rather than attempt to correct to a vocabulary of correctly spelled words, we constructed an embedding with a vocabulary that contains both correct and incorrectly spelled words. We do this by using standard algorithms \cite{Word2Vec} on a large corpus of student responses (approximately 160 million responses). The embedding we created reflects the imperfect manner in which students use words \cite{mispellingEmbedding}. For example, while the words 'happems' and 'ocures' are both incorrectly spelled versions of 'happens' and 'occurs' respectively, our embedding exhibits a high cosine similarity between the word vectors of the correct and incorrect versions. The embedding we created was an embedding into 200 dimensional space with a vocabulary consisting of 1.12 million words. Using spelling dictionaries we approximate that the percentage of correctly spelled words in the vocabulary of this embedding is approximately 7\%, or roughly 80,000 words, while the remaining 93\% are either misspellings, made up words or words from other languages. Lastly, due to the prevalence of words that are concatenated (due to a missing space), we split up any word with a Levenstein distance that is greater than two from our vocabulary into smaller words that are in the vocabulary. This ensures that any sentence is tokenized into a list of elements, almost all of which have valid embeddings.

In our classification of alerts, with respect to how they are identified by the team of reviewers, we have two tiers of alerts, Tier A and Tier B. Tier A consists of true responses that are alarming and require urgent attention while Tier B consists of responses that are concerning in nature but require further review. For simplification, both types of responses are flagged as alerts are treated equivalently by the system. This means the classification we seek is binary. Table \ref{TierA} and Table \ref{TierB} outline certain subcategories of this classification in addition to some example responses.

\begin{table}[!ht]
\ra{1.3}
\begin{tabular}{p{3cm}  p{5cm}  p{8cm}}\\ \toprule
{\bf Category A} & {\bf Details} & {\bf Examples} \\ \toprule
Harm to self or another being. & Suicide, self-harm, or extreme depression; threats or reports of violence, rape, abuse, drugs, or neglect. & I wanna kill myself. Why does my dad beat me at night?\\  \midrule
Contains mention of a gun.  & Doesn’t have to be threatening. & I want a sniper rifle.\\ \midrule
Specific and serious request for help.& Not test-related &  I hate my life, please help.\\\bottomrule\\
\end{tabular}
\caption{Student response that meet the requirement to trigger an immediate alert notification to client. \label{TierA}}
\end{table}

\begin{table}[!ht]
\ra{1.3}

\begin{tabular}{p{3cm}   p{5cm}  p{8cm}}\toprule
{\bf Category B} &  {\bf Details} & {\bf Examples} \\ \toprule
Non-specific request for help. & Not specific to test or harm.& I need help real bad help me please.\\  \midrule
Sexual Imagery. & Without threats or reports of abuse. & My uncle touches me.\\  \midrule
Violent words or phrases. & No explicit reports of being the perpetrator or victim of violence, but text seems suspect.& Death by suffocation.\\ \midrule
Signs of depression, self-loathing, or anxiety .
& Sad, lack of social support, dissatisfaction for life, grief, anxiety, negative attitude towards self. & No one loves me. I am stupid and ugly.\\\bottomrule \\
\end{tabular}
\caption{Student response contains watch words or phrases (these may be required to undergo review further). \label{TierB}}
\end{table}

The American Institutes for Research has a hand-scoring team specifically devoted to verifying whether a given response satisfies the requirements of being an alert. At the beginning of this program, we had very few examples of student responses that satisfied the above requirements, moreover, given the diverse nature of what constitutes an alert, the alerts we did have did not span all the types of responses we considered to be worthy of attention. As part of the initial data collection, we accumulated synthetic responses from the sites Reddit and Teen Line that were likely to be of interest. These were sent to the hand-scoring team and assessed as if they were student responses. The responses pulled consisted of posts from forums that we suspected of containing alerts as well as generic forums so that the engine produced did not simply classify forum posts from student responses. We observed that the manner in which the students engaged with the our essay platform in cases of alerts mimicked the way in which students used online forums in a sufficiently similar manner for the data to faithfully represent real alerts. This additional data also provided crucial examples of classes of alerts found too infrequently in student data for a valid classification. This initial data allowed us to build preliminary models and hence build better engines. 

Since the programs inception, we have greatly expanded our collection of training data, which is summarized below in Table \ref{traindat}. While we have accumulated over 1.11  million essay responses, which include many types of essays over a range of essay topics, student age ranges, styles of writing as well as a multitude of types of alerts, we find that many of them are mapped to the same set of words after applying our preprocessing steps. When we disregard duplicate responses after preprocessing, our training sample consists of only 866,137 unique responses. 

\begin{table}[!ht]
\begin{tabular}{l  l | c c c c} \toprule
Category & & Alerts & Normal & Unclassified & Total \\ \toprule
Training & Synthetic & 5012 & 67025&0& 72037 \\
(with Duplicates)& Real &7448  & 1035530&0 & 1042978\\ \midrule
Total&& 12460 & 1102555 & 0 & 1115015 \\ \toprule
Training & Synthetic &4912&57988& 0& 62900 \\ 
(Unique) & Real &5615 &797622& 0 & 803237\\ \midrule
Total && 10527 & 855610 & 0 & 866137 \\ \toprule
Threshold data & & 14 & 186 &199814 & 200014 \\ \bottomrule \\
\end{tabular}
\caption{The table gives the precise number of examples used in training both before preprocessing (with possible duplicates) and after (unique responses) as well as an unclassified set we used for determining an approximation of the percentage of responses flagged by the engine at various levels of sensitivity.\label{traindat}}
\end{table}

 Our training sample has vastly over-sampled alerts compared with a typical responses in order to make it easier to train an engine. This also means that a typical test train split would not necessarily be useful in determining the efficacy of our models. The metric we use to evaluate the efficacy of our model is an approximation of the probability that a held-out alert is flagged if a fixed percentage of a typical population were to be flagged as potential alerts. 
 
This method also lends itself to a method of approximating the number of alerts in a typical population. we use any engine produced to score a set of responses, which we call the threshold data, which consisted of a representative sample of 200,014 responses. Using these scores and given a percentage of responses we wish to flag for review, we produce a threshold value in which scores above this threshold level are considered alerts and those below are normal responses. This threshold data was scored using our best engine and the 200 responses that looked most like alerts were sent to be evaluated by our hand-scorers and while only 14 were found to be true alerts. Using the effectiveness of the model used, this suggests between 15 and 17 alerts may be in the entire threshold data set. We aggregated the estimates at various levels of sensitivity in combination with the efficacy of our best model to estimate that the rate of alerts is approximately 77 to 90 alerts per million responses. Further study is required to approximate what percentage are Tier A and Tier B.

\section{Recurrent Structures Considered}\label{RNNs}

Since natural languages contain so many rules, it is inconceivable that we could simply list all possible combinations of words that would constitute an alert. This means that the only feasible models we create are statistical in nature. Just as mathematicians use elementary functions like polynomials or periodic functions to approximate smooth functions, recurrent neural networks are used to fit classes of sequences. Character-level language models are typically useful in predicting text \cite{charlevel}, speech recognition \cite{Speech} and correcting spelling, in contrast it is generally accepted that semantic details are encoded by word-embedding based language models \cite{embedNLP}. 

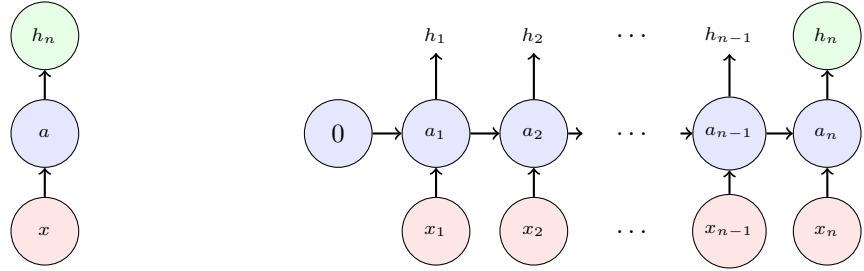
\begin{figure}[!ht]
\begin{tikzpicture}[scale=1.3]
\node [draw, fill=blue!10,circle,minimum size=.9cm] (a) at (-3,1) {${}_a$};
\node [draw, fill=blue!10,circle,minimum size=.9cm] (0) at (0,1) {$0$};
\node [draw, fill=blue!10,circle,minimum size=.9cm] (a1) at (1,1) {${}_{a_1}$};
\node [draw, fill=blue!10,circle,minimum size=.9cm] (a2) at (2,1) {${}_{a_2}$};
\node [draw, fill=blue!10,circle,minimum size=.9cm] (anm1) at (4,1) {${}_{a_{n-1}}$};
\node [draw, fill=blue!10,circle,minimum size=.9cm] (an) at (5,1) {${}_{a_n}$};
\node at (3,1) {$\ldots$};
\node at (3,0) {$\ldots$};
\node at (3,2) {$\ldots$};
\node [draw, fill=red!10,circle,minimum size=.9cm] (x) at (-3,0) {${}_x$};
\node [draw, fill=red!10,circle,minimum size=.9cm] (x1) at (1,0) {${}_{x_1}$};
\node [draw, fill=red!10,circle,minimum size=.9cm] (x2)at (2,0) {${}_{x_2}$};
\node [draw, fill=red!10,circle,minimum size=.9cm] (xnm1) at (4,0) {${}_{x_{n-1}}$};
\node [draw, fill=red!10,circle,minimum size=.9cm] (xn) at (5,0) {${}_{x_n}$};

\node [draw, fill=green!10,circle,minimum size=.9cm] (h) at (-3,2) {${}_{h_n}$};
\node (h1) at (1,2) {${}_{h_1}$};
\node (h2)at (2,2) {${}_{h_2}$};
\node (hnm1) at (4,2) {${}_{h_{n-1}}$};
\node [draw, fill=green!10,circle,minimum size=.9cm] (hn) at (5,2) {${}_{h_n}$};

\draw[->,thick] (x)--(a);
\draw[->,thick] (x1)--(a1);
\draw[->,thick] (x2)--(a2);
\draw[->,thick] (xnm1)--(anm1);
\draw[->,thick] (xn)--(an);
\draw[->,thick] (0)--(a1);
\draw[->,thick] (a1)--(a2);
\draw[->,thick] (a2) -- (2.5,1);
\draw[->,thick] (3.5,1) -- (anm1);
\draw[->,thick] (anm1) -- (an);
\draw[->,thick] (a1) -- (h1);
\draw[->,thick] (a2) -- (h2);
\draw[->,thick] (anm1) -- (hnm1);
\draw[->,thick] (an) -- (hn);
\draw[->,thick] (a) -- (h);
\end{tikzpicture}
\caption{\label{RNN}When we unfold an RNN, we express it as a sequence of cell each accepting, as input, an element of the sequence. The output of the RNN is the output of the last state.}
\end{figure}

Recurrent neural networks are behind many of the most recent advances in NLP. We have depicted the general structure of an unfolded recurrent unit in figure \ref{RNN}. A single unit takes a sequence of inputs, denoted $x$ below, which affects a set of internal states of the node, denoted $a$, to produce an output, $h_n$. A single unit either outputs a single variable, which is the output of the last node, or a sequence of the same length of the input sequence, $h = (h_1, \ldots, h_n)$, which may be used as the input into another recurrent unit. 

A layer of these recurrent units is a collection of independent units, each of which may pick up a different aspect of the series. A recurrent layer, consisting of $t$ independent recurrent units, has the ability to take the most important/prevalent features and summarize those features in a vector of length $t$. When we feed the sequence of outputs of one recurrent layer into another recurrent layer, we call this a stacked recurrent layer. Analogous to the types of features observed in stacking convolutional and dense layers in convolutional neural networks \cite{DeepImages}, it is suspected that stacking recurrent layers allows a neural network to model more semantically complex features of a text \cite{DeepRNN, DeepOpinionRNN}.

The collections of variables associated with the state of the recurrent units, which are denoted $a$ in figure \ref{RNN}, and their relations between the inputs, $x$, and the outputs are what distinguishes simple recurrent units, GRUs and LSTM units. In our case, $x =(x_1,x_2, \ldots, x_n)$ is a sequence of word-vectors. The underlying formulas for gated recurrent units are specified by the initial condition $h_0 = 0$ and  
\begin{subequations}
\begin{align}
z_t &= \sigma_g (W_z x_t + U_z h_{t-1} + b_z),\\
r_t &= \sigma_g (W_r x_t + U_r h_{t-1} + b_r),\\
h_t & = \tilde{z}_t \circ h_{t-1} + z_t \circ y_t,\\
y_t &= \sigma_h (W_h x_t + U_h(r_t \circ h_{t-1}) + b_h),
\end{align}
\end{subequations}
where $\circ$ denotes the element-wise product (also known as the Hadamard product), $x_t$ is an input vector $h_t$ is an output vector, $z_t$ is and update gate, $\tilde{z}_t = 1-z_t$, $r_t$ is a reset gate, subscripted variables $W$, $U$ and $b$ are parameter matrices and a vector and $\sigma_h$ and $\sigma_h$ are the original sigmoid function and hyperbolic tangent functions respectively \cite{GRU2014}.

The second type of recurrent unit we consider is the LSTM, which appeared in the literature before the GRU and contains more parameters \cite{LSTM1997}. It was created to address the vanishing gradient problem and differs from the gated recurrent unit in that it has more parameters, hence, may be regarded as more powerful. 
\begin{subequations}
\begin{align}
f_t &= \sigma_g (W_f x_t + U_f h_{t-1} + b_f),\\
i_t &= \sigma_g (W_i x_t + U_i h_{t-1} + b_i),\\
o_t &= \sigma_g (W_o x_t + U_o h_{t-1} + b_o),\\
c_t &= f_t \circ c_{t-1} + i_t \circ y_t,\\
h_t &= o_t \circ \sigma_h(c_t), \\
y_t & =  \sigma_h (W_z x_t + U_z h_{t-1} + b_z),
\end{align}
\end{subequations}
where $x_t$ is the input, $c_t$ is the cell state vector, $f_t$ is the forget gate, $i_t$ is the input gate, $o_t$ is the output gate and $h_t$ is the output, $z_t$ is a function of the input and previous output while subscripted variables $W$, $U$ and $b$ are parameter matrices and a vector. Due to their power, LSTM layers are ubiquitous when dealing with NLP tasks and are being used in many more contexts than layers of GRUs \cite{NLPLSTM}. 
\begin{figure}[!ht]
\begin{tikzpicture}[scale=.9]
\draw[fill=blue!10] (0,0) rectangle (6,3);
\node [draw, fill=green!10,circle,minimum size=1cm] (ct) at (4,1.5) {${}_{c_t}$};
\draw[very thick,->] (ct) -- (4,3.5);
\draw[very thick,->] (-.5,1.5) -- (ct);
\node [draw, fill=red!10,rectangle,minimum size=.5cm] (ft) at (.7,1.5) {${}_{f_t}$};
\node [draw, fill=red!10,rectangle,minimum size=.5cm] (ot) at (4,2.5) {${}_{o_t}$};
\node [draw, fill=red!10,rectangle,minimum size=.5cm] (it) at (3,.7) {${}_{i_t}$};
\node [draw, fill=yellow!10,circle] (yt) at (1.4,.7) {${}_{y_t}$};
\draw[very thick] (it) -- (yt);
\node[draw, fill=blue!10,circle,minimum size=.2cm] (plus) at (3,1.5) {};
\node at (3,1.5) {$+$};
\draw[very thick] (plus) -- (it);
\node at (-.8,1.5) {$c_{t-1}$};
\node at (4,3.8) {$h_{t}$};
\node at (6.5,2.5) {$h_{t}$};
\node at (1.4,-.5) (xt) {$x_{t}$};
\node at (-.8,.7) (htm1) {$h_{t-1}$};
\draw[very thick,->] (xt) -- (yt);
\draw[very thick,->] (ot) -- (6.2,2.5);
\draw[very thick,->] (htm1) -- (yt);
\draw[very thick,->] (ct) -- (6.2,1.5);
\node at (6.5,1.5) {$c_t$};
\end{tikzpicture}
\hspace{1cm}
\begin{tikzpicture}[scale=.9]
\draw[fill=blue!10] (0,0) rectangle (6,3);
\node at (3.8,3.8) (ht) {$h_{t}$};
\node at (4.6,-.5) (xt) {$x_{t}$};
\node at (-.8,.7) (htm1) {$h_{t-1}$};
\node [draw, fill=red!10,rectangle,minimum size=.5cm] (zt) at (4.6,1.8) {${}_{z_t}$};
\node [draw, fill=yellow!10,circle] (yt) at (4.6,.7) {${}_{y_t}$};
\draw[very thick,->] (-.4,.7) --(yt);
\draw[very thick,->] (xt) --(yt);
\draw[very thick] (yt) --(zt);
\draw[very thick] (1.5,.7) --(1.5,1.8) -- (zt);
\node[draw, fill=blue!10,circle,minimum size=.2cm] (plus) at  (3.8,1.8) {};
\node at (3.8,1.8) {$+$};
\node [draw, fill=red!10,rectangle,minimum size=.5cm] (tzt) at (3,1.8) {${}_{\tilde{z}_t}$};
\node [draw, fill=red!10,rectangle,minimum size=.5cm] (ot) at (3,.7) {${}_{r_t}$};
\draw[very thick,->] (plus) --(ht);
\draw[very thick,->] (3.8,2.5) --(6.2,2.5);
\node at (6.5,2.5) {$h_{t}$};
\end{tikzpicture}
\caption{The left is an LSTM; the gates $f_t$, $i_t$ and $o_t$ are vectors of values between 0 and 1 that augment the input being passed through them by selecting which features to keep and which to discard. The input $x_t$, $h_{t-1}$, $c_{t-1}$ and $y_t$ are more generically valued vectors. The GRU is based on a similar concept with a simpler design where an update gate, $z_t$ decides which features of the previous output to keep as output of the new cell and which features need to contain input specific information, which is stored in $y_t$.}
\end{figure}
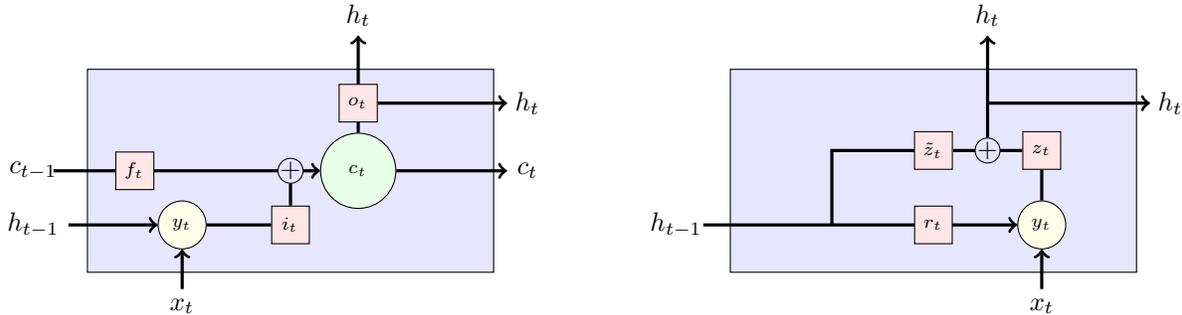

Given a recurrent unit, the sequence $x$ is fed into the recurrent unit cell by cell in the order it appears, however, it was found that some recurrent networks applied to translation benefited from reversing the ordering of the sequence, so that the recurrent units are fed the vectors from last to first as opposed to first to last. Indeed, it is possible to state the most important information at the beginning of a text or at the end. The idea behind bidirectional recurrent units is that we double the number of set units and have half the units fed the sequence in the right order, while the other half of the units are fed the sequence in reverse. Due to the lack of symmetry in the relations between states, we are potentially able to model new types of sequences in this way.

The last mechanism we wish to test is an attention mechanism \cite{Bahdanau:Attention}. The key to attention mechanisms is that we apply weights to the sequences, $h_t$, outputted by the recurrent layer, not just the final output. This means that the attention is a function of the intermediate states of the recurrent layer as well as the final output. This may be useful when identifying when key phrases are mentioned for example. This weighted sequence is sent to a soft-max layer to create a context vector. The attention vector is then multiplied by $h_t$ to produce resulting attention vector, $\tilde{h}_t$. We have implemented the following attention mechanism
\begin{subequations}
\begin{align}
\tilde{h}_t &= c_t h_t,\\
c_t &= \sum \alpha_{t,j} h_j,\\
\alpha_{i,j} &= \dfrac{\exp(e_{ij})}{\sum \exp(e_{ik})},
\end{align}
\end{subequations}
where $h_t$ was the output from the LSTM layer, the $e_{ij}$ are linear transformations of the $h_j$ and $\tilde{h}_t$ is the attended output, i.e., the output of the attention layer \footnote{An attention function, e.g., $f = \mathrm{tanh}$, is often used so that $\tilde{h}_t = f(c_t h_t)$. We tested this approach only to find similar overall results in our experiments \cite{MPM2015:Attention}.}. This mechanism has been wildly successful in machine translation \cite{MPM2015:Attention, GoogleTranslation} and other tasks \cite{Xu}. 

\section{Methodology and Results}\label{sec:Results}

Unlike many tasks in NLP, our goal is not to explicitly maximize accuracy. The framework is that we may only review a certain percentage of documents, given this, we want to maximize the probability than an alert will be caught. I.e., the cost of a false-positive is negligible, while we consider false negatives to be more serious. Conversely, this same information could be used to set a percentage of documents required to be read in order to have have some degree of certainty that an alert is flagged. If we encode all alerts with the value 1 and all normal documents with a value of 0, any neural network model will serve as a statistical mechanism in which an alert that was not used in training will, a priori, be given a score by the engine from a distribution of numbers between 0 and 1 which is skewed towards 1 while normal documents will also have scores from another distribution skewed towards 0. The thresholds values where we set are values in which all scores given by the engine above the cut-off are considered possible alerts while all below are considered normal. We can adjust the number of documents read, or the percentage of alerts caught by increasing or decreasing this cut-off value.

To examine the efficacy of each model, our methodology consisted of constructing three sets of data:
\begin{enumerate}
\item{A small collection of alerts that was removed from training.}
\item{Our training set consisting of almost all our data except for a small collection of alerts.}
\item{A sufficiently large corpus of unclassified generic test responses, which was sampled as uniformly as possible over the various sources of responses.}
\end{enumerate}
The idea is that we use the generic test responses to determine how each model would score the types of responses the engine would typically see. While the number of alerts in any set can vary wildly, it is assumed that the set includes both normal and alert responses in the proportions we expect in production. Our baseline model is logistic regression applied to a TF-IDF model with latent semantic analysis used to reduce the representations of words to three hundred dimensions. This baseline model performs poorly at lower thresholds and fairly well at higher thresholds.

\begin{table}[!ht]
\begin{tabular}{l c c c c c  c  c }\toprule
{\bf Model} & {\bf Configuration} & {\bf 0.1 \%} & {\bf 0.3\%} & {\bf 0.5\%} & {\bf 1\%} & {\bf 2\%} & {\bf 4\%} \\ \toprule
BOW + LSA + Logistic Regression&  & 64.8 & 82.2 & 86.9 & 91.3 &  94.6 & 96.3 \\\toprule
GRU & (512) & 51.4 & 62.3 & 67.4 & 73.9 & 80.3 & 85.4\\ \midrule
Stacked GRU & (256,256) & 73.2 & 80.6 & 83.5 & 87.1 & 90.6 & 93.2\\ \midrule
Bidirectional GRU & (256) & 55.7 & 64.7 & 68.8 & 74.9 & 81.6 & 86.6\\ \midrule
Bidirectional Stacked GRU & (128,128) &  80 & 86.8 & 89.2 & 92.2 & 94.4 & 95.2\\ \midrule
GRU with Attention & (512) & 55.4 & 67.5 & 72 & 77.6 & 84 & 91.9\\ \midrule
Stacked GRU with Attention & (256,256) & 69.1 & 78.7 & 81.9 & 86.2 & 89.5 & 92.6 \\ \midrule
Bidirectional GRU with Attention& (256) &59.8 & 70 & 75.6 & 80.6 & 87.4 & 93.3\\ \midrule
Bidirectional Stacked GRU with Attention & (128,128) & 76.6 & 84.2 & 85.9& 90.2 & 93.4 & 95.3\\ \midrule
LSTM &(512) & 66.6 & 72.8 & 75.8 & 78.6 & 86 & 92.1 \\ \midrule
Stacked LSTM &(256,256) & 80.8 & 87.5 &89.6 & 93 & 94.8 & 96.7 \\ \midrule
Bidirectional LSTM &(256) & 62 & 69.3  & 72.6 & 77.8 & 83 & 87.5\\ \midrule
Bidirectional Stacked LSTM & (128,128) & 83.5 & 87.8 & 90.6 & 93.2 & 94.2 & 96.2 \\ \midrule
LSTM with Attention & (512) & 62.2 & 79.4 & 85.6 & 89.5 & 92.8 & 95.6\\ \midrule
Stacked LSTM with Attention &(256,256) & 86 & 90.3  & 91.7  & 93.6 & 95.3 & 97.2 \\ \midrule
Bidirectional LSTM with Attention &(256) & 66.5 & 81.5 & 86.4 & 91.5 & 94.3 & 96.4 \\ \midrule
Bidirectional Stacked LSTM with Attention & (128,128) & {\bf 86.2} & {\bf 91} & {\bf 93.5} & {\bf 95.5} & {\bf 96.8} & {\bf 98.7}\\ \bottomrule \\
\end{tabular}
\caption{Approximations of the percentage of alerts caught by each model for each percentage allowed to be reviewed. \label{Results}}
\end{table}

To evaluate our models, we did a 5-fold validation on a withheld set of 1000 alerts. That is to say we split our set into 5 partitions of 200 alerts, each of which was used as a validation sample for a neural network trained on all remaining data. This produced five very similar models whose performance is given by the percentage of 1000 alerts that were flagged. The percentage of 1000 alerts flagged was computed for each level of sensitivity considered, as measured by the percentage of the total population flagged for potentially being an alert.

Each of the models had 512 recurrent units (the attention mechanisms were not recurrent), hence, in stacking and using bidirectional variants, the number of units were halved. We predominantly trained on using Keras with Tensorflow serving the back-end. The machines we used had NVIDIA Tesla K80s. Each epoch took approximately two to three hours, however, the rate of convergence was such that we could restrict our attention to the models formed in the first 20 epochs as it was clear that the metrics we assessed had converged fairly quickly given the volume of data we had. The total amount of GPU time spent on developing these models was in excess of 4000 hours. 

To give an approximation of the effect of each of the attributes we endowed our models with, we can average over the effectiveness of each model with and without each attribute in question. It is clear that that stacking two layers of recurrent units, each with half as many cells, offers the greatest boost in effectiveness, followed by the difference in recurrent structures followed by the use of attention. Using bidirectional units seems to give the smallest increase, but given the circumstances, any positive increase could potentially save lives. 

\begin{table}[!ht]
    \centering
    \begin{tabular}{l|c}\toprule
     {\bf Attribute} & Effect \\\toprule
     LSTM vs GRU & 8.05\% increase \\ 
     Bidirectional vs Normal & 2.20\% increase \\
     Attention vs No Attention & 4.15\% increase \\
     Stacked vs Flat & 14.98 \% increase \\ \bottomrule 
    \end{tabular}
    \caption{The effect of each of the attributes we endowed our networks.}
    \label{attributes}
\end{table}

\section{Conclusions}\label{Conclusion}

The problem of depression and violence in our schools is one that has recently garnered high levels of media attention. This type of problem is not confined to the scope of educational research, but this type of anomaly detection is also applicable to social media platforms where there are posts that indicate potential cases of users alluding to suicide, depression, using hate-speech and engaging in cyberbullying. The program on which this study concerns is in place and has contributed to the detection an intervention of cases of depression and violence across America. This study itself has led to a dramatic increase in our ability to detect such cases.

We should also mention that the above results do not represent the state-of-the-art, since we were able to take simple aggregated results from the models to produce better statistics at each threshold level than our best model. This can be done in a similar manner to the work of \cite{Ensemble}, however, this is a topic we leave for a future paper. It is also unclear as to whether traditional sentiment analysis provides additional information from which better estimates may be possible.  

\section{Acknowledgements}

I would like to thank Jon Cohen, Amy Burkhardt, Balaji Kodeswaran, Sue Lottridge and Paul van Wamelen for their support and discussions.

\end{document}